\RequirePackage{fix-cm}

\documentclass[twocolumn]{svjour3}
\usepackage{mathptmx}
\usepackage{amsmath}
\usepackage{graphicx}
\usepackage{epstopdf}
\usepackage{fixltx2e}

\smartqed
\spnewtheorem*{Remark}{Remark}{\bf}{\it}
\spnewtheorem{algorithm}{Algorithm}{\bf}{\rm}

\begin{document}

\title{Nested Dissection Solver for Transport in 3D Nano-Electronic Devices}

\author{
Y. Zhao \and
U. Hetmaniuk \and
S. R. Patil\and
J. Qi\and
M. P. Anantram}

\institute{Y. Zhao \and S. R. Patil\and J. Qi\and M. P. Anantram \at
              Department of Electrical Engineering,
                University of Washington, Box 352500,
                Seattle, WA 98195-2500, U.S.A. \\
              \email{zhaoyq@uw.edu}
           \and
           U. Hetmaniuk \at
             Department of Applied Mathematics,
                University of Washington, Box 353925,
                Seattle, WA 98195-3925, U.S.A.
}

\date{Received: \today}

\maketitle
\begin{abstract}
The Hierarchical Schur Complement method (HSC), and the HSC-extension, have significantly accelerated the evaluation of the retarded Green's function, particularly
the lesser Green's function, for two-dimensional nanoscale devices.
In this work, the HSC-extension is applied to determine the solution of non-equilibrium Green's functions (NEGF) on three-dimensional nanoscale devices.
The operation count for the HSC-extension is analyzed for a cuboid device.
When a cubic device is discretized with $N \times N \times N$ grid points, the state-of-the-art Recursive Green Function (RGF) algorithm takes $\mathcal{O}(N^7)$ operations, whereas the HSC-extension only requires $\mathcal{O}(N^6)$ operations.
Operation counts and runtimes are also studied for three-dimensional nanoscale devices of practical interest: a graphene-boron\- nitride-graphene multilayer system, a silicon nanowire, and a DNA molecule.
The numerical experiments indicate that the cost for the HSC-extension is proportional to the solution of one linear system (or one LU-factorization) and that the runtime speed-ups over RGF exceed three orders of magnitude when simulating realistic devices, such as a graphene-boron nitride-graphene multilayer system with 40,000 atoms.

\end{abstract}

\section{Introduction}

With the downscaling of nanoscale electronic devices, the non-equilibrium Green's function (NEGF) method has become the most effective and accurate modeling approach in predicting the electronic transport performance.
The self-consistent solution of NEGF and Poisson's equations is capable of  accurately modeling coherent as well as decoherent transport by integrating atomic/molecular orbital level physics \cite{datta2002non}. An aspect of NEGF, that currently prevents a broader use, is the large computational cost associated with the method.

Researchers and analysts have used several approximations to alleviate this cost.
Examples include the mode-space approximation \cite{ren2003nanomos},
which couples 1D NEGF transport simulation to transverse states from solving
2D Schr\"{o}dinger equation on the cross-section, and the restriction to levels of lower accuracy (such as tight-binding or effective mass levels).
Although these approximations of NEGF have successfully predicted the transport characteristics of nanoscale devices \cite{barker1989theory,luisier2006quantum}, these alternatives remain unable to capture atomic-scale inhomogeneities, such as surface roughness, unintentional doping, and trapped charges (see, for example, \cite{asenov2003simulation} for an illustration of the resulting inaccuracy). The modeling of such inhomogeneities requires a full 3D real-space NEGF simulation.
Recent studies \cite{martinez2007study,martinez2006development,martinez2007self,martinez2009comparison,martinez2010detailed} have performed 3D NEGF simulations to handle these inhomogeneities but with ``coarse discretization'' ({\it i.e.} small number of atomic orbitals or small number of grid points) to control the computational cost.
To enhance the predictability of these 3D NEGF simulations, finer discretizations have to be considered.
Therefore the computational cost of NEGF needs to be addressed.
The goal of this paper is to present a numerical method that  significantly reduces the cost of 3D NEGF simulations.

During a NEGF simulation, the main computational bottleneck is the repeated evaluations of the retarded Green's function, $G^r$, and the lesser Green's function, $G^{<}$, over a wide range of energy values \cite{Anantram2008aa}.
Physical quantities, such as density of states, carrier density,
and transmission coefficients, are obtained from the evaluations of these
Green's functions.
After discretization of the nanoscale device, these repeated evaluations
amount to computing entries in the matrices $\mathbf{G}^r$ and
$\mathbf{G}^{<}$ that approximate, respectively, the retarded and lesser
Green's functions.
Recall that the matrix $\mathbf{G}^{r}$ is defined as
\begin{equation}
\mathbf{G}^{r} (E) = \mathbf{A}(E) ^{-1}
\quad \mbox{ with } \quad
\mathbf{A}(E) = E \mathbf{I} - \mathbf{H} - \boldsymbol{\Sigma}^{r}
  \label{eq:gr}
\end{equation}
where $\mathbf{H}$ is the system Hamiltonian matrix and $\mathbf{I}$ is the identity matrix.
The self-energy matrix $\boldsymbol{\Sigma}^{r}$ consists of two parts: the contact self-energy $\boldsymbol{\Sigma}^{r}_{C}$ and a self-energy to represent scattering mechanisms $\boldsymbol{\Sigma}^{r}_{scant}$.
The matrix $\mathbf{G}^{<}$ is defined as
\begin{equation}
\mathbf{G}^{<} (E) =  \mathbf{G}^{r} (E) \boldsymbol{\Sigma}^{<} (E) \left( \mathbf{G}^{r} (E) \right)^{\dagger}
 \label{eq:gl}
\end{equation}
where $\boldsymbol{\Sigma}^{<}$ corresponds to the lesser self-energy matrix \cite{Anantram2008aa}.
To compute diagonal and desired off-diagonal entries of $\mathbf{G}^{r}$ and $\mathbf{G}^{<}$,
the recursive Green's function algorithm (RGF) \cite{Svizhenko2002aa}
has been the algorithm of choice for many years.
Recently, the hierarchical Schur complement (HSC) method \cite{Lin2009ab,Lin2011ac} and the fast inverse using nested dissection (FIND) method \cite{Li2008ab,Li2011ab} have exhibited significant speed-ups over RGF to evaluate the diagonal entries of $\mathbf{G}^{r}$.
Both of these methods have a smaller operation count than the RGF method.
Our HSC-extension to compute diagonal and desired off-diagonal entries of $\mathbf{G}^{<}$ \cite{hetmaniuk2013nested} has also demonstrated significant speed-ups over RGF simulations for 2D nanoscale devices.
In this paper, the efficiency of the HSC-extension \cite{hetmaniuk2013nested} over the RGF method is studied for 3D nanoscale devices.

The rest of the paper is organized as follows.
Section~\ref{sec:algorithm} gives a brief description of the RGF and HSC-extension approaches and discusses the asymptotic operation counts for both approaches when simulating 3D devices.
Section~\ref{sec:results} presents the runtime analysis for a cuboid structure with a regular mesh and three state-of-the-art nanoscale devices: a graphene-boron nitride device with a multilayer hexagonal mesh; a silicon nanowire structure with 10 localized atomic orbitals; and a DNA molecule sequence with an unstructured atomic connectivity.
Section~\ref{sec:conc} provides a brief summary of the numerical experiments.

\section{Algorithms overview and cost analysis}
\label{sec:algorithm}

This section briefly describes the RGF method \cite{Svizhenko2002aa} and
the HSC-extension approach \cite{hetmaniuk2013nested}.
Then the operation counts for both methods are discussed when simulating
cuboid or brick-like nanoscale devices.

\subsection{Overview of the RGF algorithm}

The RGF algorithm \cite{Svizhenko2002aa} evaluates the specific entries
of the matrices $\mathbf{G}^{r}$ and $\mathbf{G}^{<}$ when the system
matrix $\mathbf{A}$ is structured as a block-tridiagonal matrix.
The off-diagonal and diagonal blocks of $\mathbf{A}$ are uniquely
associated with layers (of atoms or of grid points) orthogonal
to the transport direction, usually the $y$-direction.
The RGF algorithm is composed of two passes to compute diagonal entries
of $\mathbf{G}^{r}$ and two passes to compute diagonal and desired off-diagonal entries
of $\mathbf{G}^{<}$.
The two passes to compute entries of $\mathbf{G}^{r}$ (or of $\mathbf{G}^{<}$)
are interpreted as follows:
\begin{itemize}
\item
the first pass marches one layer at a time from {\it left to right} along
the $y$-direction and, recursively, {\it folds} the effect of left layers
into the current layer;
\item
the second pass marches one layer at a time from {\it right to left} along
the $y$-direction and, recursively, {\it extracts} the diagonal blocks and
the nearest neighbor off-diagonal blocks for the final result.
\end{itemize}

Mathematically, when computing diagonal entries of $\mathbf{G}^{r}$,
the first pass amounts to transforming $\mathbf{A}$ into a block diagonal
matrix $\mathbf{D}$,
\begin{equation}
\mathbf{L}^{-1} \mathbf{A} \mathbf{L}^{-T}
=
\mathbf{D}
,
\end{equation}
where $\mathbf{L}$ is a lower block-triangular matrix with identity blo\-cks
on the diagonal.
Recall that $\mathbf{A}$ being a complex symmetric matrix, the block
$\mathrm{LDL}^T$-factorization is a particular form of a block
LU-factorization.
The second pass in the evaluation of entries for $\mathbf{G}^{r}$ is
expressed by the formula
\begin{equation}
\label{eqn:takahashi_gr}
\mathbf{G}^{r}
=
\left( \mathbf{I} - \mathbf{L}^T \right) \mathbf{G}^{r}
+
\mathbf{D}^{-1} \mathbf{L}^{-1}
\end{equation}
(see \cite{Takahashi1973,Erisman1975} for further details).
Starting from the knowledge of the block on the diagonal of $\mathbf{G}^{r}$
for the rightmost layer\footnote{The block on the diagonal of $\mathbf{G}^{r}$
for the rightmost layer is the corresponding block
in the matrix $\mathbf{D}^{-1}$.}, the formula (\ref{eqn:takahashi_gr})
provides, one layer at a time, the diagonal block and the nearest neighbor
off-diagonal blocks when marching from {\it right to left} along
the $y$-direction.

When computing $\mathbf{G}^{<}$,
the first pass amounts to evaluating the matrix-matrix products,
\begin{equation}
\mathbf{D}^{-1}\mathbf{L}^{-1} \boldsymbol{\Sigma}^{<}
\mathbf{L}^{-\dagger} \mathbf{D}^{-\dagger}
\end{equation}
where the matrices $\mathbf{L}$ and $\mathbf{D}$ form the block
$\mathrm{LDL}^T$-factori\-za\-tion of $\mathbf{A}$.
The second pass in the evaluation of entries for $\mathbf{G}^{<}$ is
expressed by the formula
\begin{equation}
\label{eqn:takahashi_gless}
\mathbf{G}^{<}
=
\left( \mathbf{I} - \mathbf{L}^T \right) \mathbf{G}^{<}
+
\mathbf{D}^{-1} \mathbf{L}^{-1} \boldsymbol{\Sigma}^{<}
\mathbf{L}^{-\dagger} \mathbf{D}^{-\dagger}
\left( \mathbf{L}^{T} \right)^{-\dagger}
\end{equation}
(see \cite{Takahashi1973} for further details).
Starting from the knowledge of the block on the diagonal of $\mathbf{G}^{<}$
for the rightmost layer, the formula (\ref{eqn:takahashi_gless}) provides,
one layer at a time, the diagonal block and the nearest neighbor
off-diagonal blocks when marching from {\it right to left} along
the $y$-direction.

\subsection{Overview of the HSC algorithm and its extension for $\mathbf{G}^{<}$}

The HSC method and its extension \cite{Lin2009ab,Lin2011ac,hetmaniuk2013nested} evaluate specific entries of the matrices $\mathbf{G}^r$ and $\mathbf{G}^<$ when the matrix $\mathbf{A}$ is symmetric and sparse. In contrast to RGF, the matrix $\mathbf{A}$ is not required to be block-tridiagonal.

The first step of the HSC method and of its extension gathers the grid points (or atoms) into arbitrarily-shaped clusters. These clusters are then organized in a binary tree. The ordering of rows of $\mathbf{A}$ according to this hierarchy of clusters arranges the non-entries of $\mathbf{A}$ into a block arrow-shaped structure. Such choice of ordering allows {\it to fold} and {\it to extract} in any physical direction when following the vertical hierarchy of the binary tree.

\begin{figure}[hbp]
\centering
\includegraphics[width=\columnwidth]{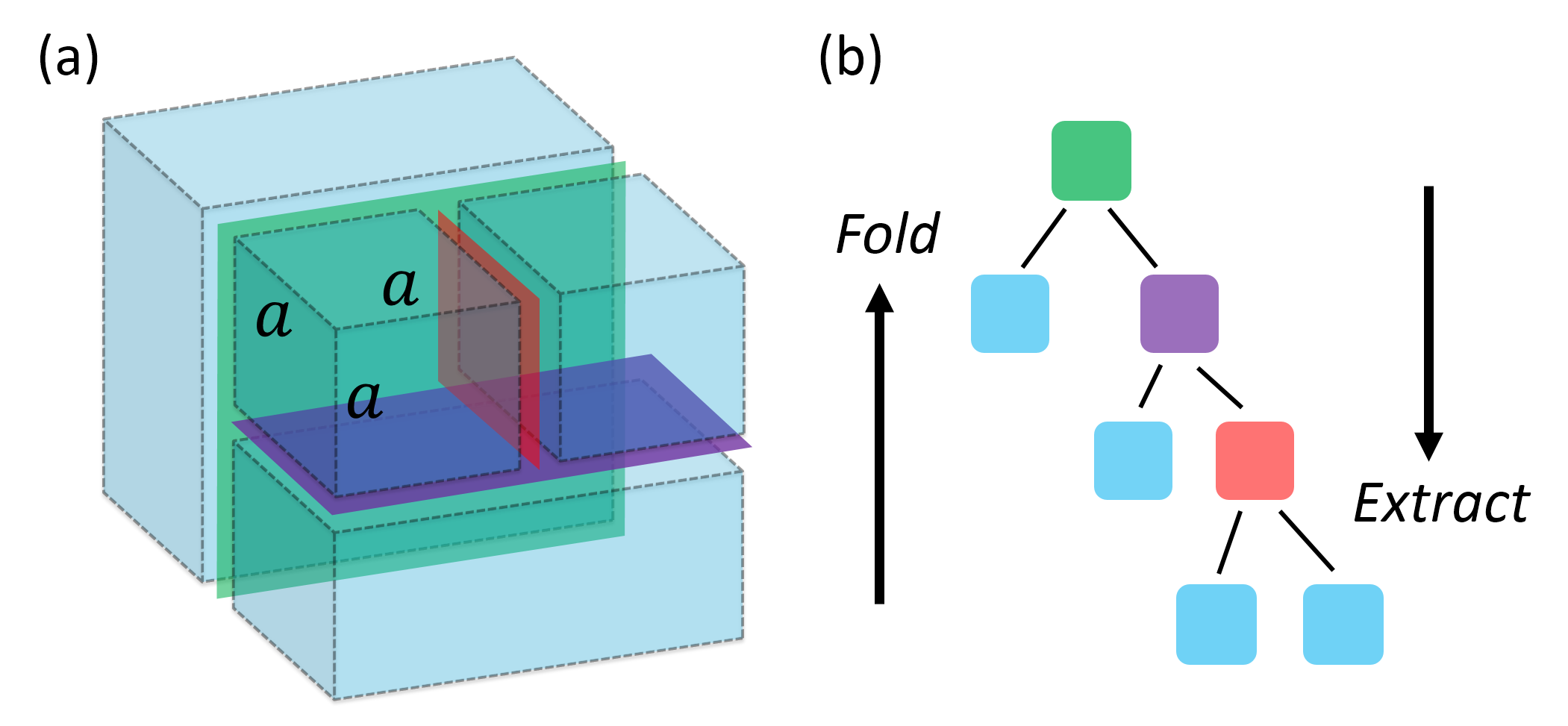}
\caption{(a) The domain decomposition from cube of dimension $2a+1$ to cubes of dimension $a$. Three levels of separators are colored by red, purple and green respectively. (b) The multilevel binary tree corresponding to the cuboid decomposition. The three levels of separators are depicted with matching colors. The blue blocks denote the corresponding blue clusters.}
\label{fig:merge}
\end{figure}

Fig.~\ref{fig:merge} illustrates a partition for a cuboid device of size $ (2a+1) \times (2a+1) \times (2a+1)$ and the corresponding binary tree\footnote{In practice, the binary tree is likely to be balanced.}.
Three levels of separators are depicted in Fig.~\ref{fig:merge} and the colors in the binary tree match the colors in the partition.
Note that the three separators are orthogonal to three distinct physical directions.
The {\it fold} process represents the computation from the lowest level to the top-most level, while the {\it extract} process marches one level at a time from the top-most level to the lowest one.

The mathematical interpretation of the two passes for $\mathbf{G}^{r}$
remains unchanged with the stipulation that the rows of $\mathbf{A}$,
$\mathbf{G}^{r}$, and $\mathbf{G}^{<}$ are ordered according to
the hierarchy of clusters in the binary tree.
The first pass for $\mathbf{G}^{<}$ is interpreted as the evaluation
of the matrix-matrix products,
\begin{equation}
\mathbf{D}^{-1}\mathbf{L}^{-1}
\boldsymbol{\Sigma}^{<} \left( \mathbf{G}^{r} \right)^{\dagger}
\end{equation}
and the second pass is expressed by the formula
\begin{equation}
\label{eqn:takahashi_gless_2}
\mathbf{G}^{<}
=
\left( \mathbf{I} - \mathbf{L}^T \right) \mathbf{G}^{<}
+
\mathbf{D}^{-1} \mathbf{L}^{-1} \boldsymbol{\Sigma}^{<}
\left( \mathbf{G}^{r} \right)^{\dagger}
\end{equation}
where the matrix rows remain ordered according to the hierarchy
of clusters in the binary tree.

Next we provide two pseudo-codes describing, respectively, the HSC method and its extension.
On the basis of the hierarchical structure of a binary tree with $L$ levels, let $P_i=\{ {\rm clusters}\ j\}$ denote the set of all cluster indices $j$ such that cluster $j$ is an ancestor of cluster $i$. Similarly, let $C_i=\{ {\rm clusters}\ j\}$ denote the set of all cluster indices $j$ such that cluster $j$ is an descendant of cluster $i$.
Algorithm~\ref{alg:gr} describes the computation of {\it specific} entries of $\mathbf{G}^{r}$ (HSC method).

\begin{algorithm}[HSC: computation of $\mathbf{G}^r$]\\
\rule[2mm]{\columnwidth}{1pt} \\
{\bf for} $l = 1$ to $L - 1$ {\bf do}\newline\indent
   $\mathbf{A}^{(l)} = \mathbf{A}^{(l-1)}$
        \hfill $\mathbf{A}^{(0)} =  \mathbf{A}$ \newline\indent
    {\bf for} all the clusters $i$ on level $l$ {\bf do}\newline\indent
        \indent $\boldsymbol{\Psi}_{i,j}=-\left(\mathbf{A}_{i,i}^{(l)}\right)^{-1}\mathbf{A}_{i,j}^{(l)}$ for all $j$ in $P_{i}$\newline\indent
        \indent $\mathbf{A}_{j,k}^{(l)}=\mathbf{A}_{j,k}^{(l)}+\boldsymbol{\Psi}_{i,j}^{T}\mathbf{A}_{i,k}^{(l)}$ for all $j$ and $k$ in $P_{i}$\newline\indent
        \indent
        $\mathbf{A}_{k,j}^{(l)}=\left(\mathbf{A}_{j,k}^{(l)}\right)^{T}$ for all $j$ and $k$ in $P_{i}$\newline\indent
        \indent
        $\mathbf{A}_{i,j}^{(l)}=\mathbf{0}$ and $\mathbf{A}_{j,i}^{(l)}=\mathbf{0}$ for all $j$ in $P_{i}$\newline\indent
    {\bf end for}\newline
{\bf end for}\\
$\mathbf{G}^{(L-1)}=\left(\mathbf{A}^{(L-1)}\right)^{-1}$
    \hfill {$\mathbf{A}^{(L-1)}$ is block diagonal}\\
{\bf for} {$l = L-2$ to $0$} {\bf do}\newline\indent
        $\mathbf{G}^{(l)}=\mathbf{G}^{(l+1)}$\newline\indent
        {\bf for} {all the clusters $i$ on level $l$} {\bf do}\newline\indent
            \indent $\mathbf{G}_{i,j}^{(l)}=\mathbf{G}_{i,j}^{(l)}+\sum_{k\in P_{i}}\boldsymbol{\Psi}_{i,k}\mathbf{G}_{k,j}^{(l)}$ for all cluster indices $j$ in $P_{i}$\newline\indent
            \indent $\mathbf{G}_{j,i}^{(l)}=\left(\mathbf{G}_{i,j}^{(l)}\right)^{T}$ for all cluster indices $j$ in $P_{i}$\newline\indent
            \indent $\mathbf{G}_{i,i}^{(l)}=\mathbf{G}_{i,i}^{(l)}+\sum_{j\in P_{i}}\boldsymbol{\Psi}_{i,j}\mathbf{G}_{j,i}^{(l)}$\newline\indent
{\bf end for}\newline
{\bf end for}\\
Computed entries in $\mathbf{G}^{(0)}$ match corresponding entries in $\mathbf{G}^r$\\
\rule[2mm]{\columnwidth}{1pt} \\
\label{alg:gr}
\end{algorithm}

Algorithm~\ref{alg:gl}  describes the computation of {\it specific} entries of $\mathbf{G}^{<}$ with the HSC-extension.
For further details about the HSC method for $\mathbf{G}^{r}$
and about its extension for $\mathbf{G}^{<}$, the reader is
referred to the references \cite{Lin2009ab,Lin2011ac,hetmaniuk2013nested}.

\begin{algorithm}[HSC-extension: computation of $\mathbf{G}^<$]\\
\rule[2mm]{\columnwidth}{1pt} \\
$\mathbf{N}=\boldsymbol{\Sigma}^{<}\left(\mathbf{G}^{(0)}\right)^{\dagger}$ 
    \hfill $\boldsymbol{\Sigma}^{<}$ block diagonal\\
{\bf for} $l = 1$ to $L - 1$ {\bf do}\newline\indent
   $\mathbf{N}^{(l)}=\mathbf{N}^{(l-1)}$
        \hfill $\mathbf{N}^{(0)} =  \mathbf{N}$ \newline\indent
    {\bf for} all the clusters $i$ on level $l$ {\bf do}\newline\indent
        \indent
        $\mathbf{N}_{j,k}^{(l)}=\mathbf{N}_{j,k}^{(l)}+\boldsymbol{\Psi}_{i,j}^{T}\mathbf{N}_{i,k}^{(l)}$ for all $j$ and $k$ in $P_{i}$ \newline\indent
    {\bf end for}\newline
{\bf end for}\\
$\mathbf{P}^{(L-1)}=\mathbf{G}^{(L-1)}\mathbf{N}^{(L-1)}$\\
{\bf for} {$l = L-2$ to $0$} {\bf do}\newline\indent
        $\mathbf{P}^{(l)}=\mathbf{P}^{(l+1)}$\newline\indent
        {\bf for} {all the clusters $i$ on level $l$} {\bf do}\newline\indent
            \indent $\mathbf{P}_{i,j}^{(l)}=\mathbf{P}_{i,j}^{(l)}+\sum_{k\in P_{i}}\boldsymbol{\Psi}_{i,k}\mathbf{P}_{k,j}^{(l)}$ for all cluster indices $j$ in $P_{i}$\newline\indent
            \indent $\mathbf{P}_{j,i}^{(l)}=-\left(\mathbf{P}_{i,j}^{(l)}\right)^{\dagger}$ for all cluster indices $j$ in $P_{i}$\newline\indent
            \indent $\mathbf{P}_{i,i}^{(l)}=\mathbf{P}_{i,i}^{(l)}+\sum_{j\in P_{i}}\boldsymbol{\Psi}_{i,j}\mathbf{P}_{j,i}^{(l)}$\newline\indent
{\bf end for}\newline
{\bf end for}\\
Computed entries in $\mathbf{P}^{(0)}$ match corresponding entries in $\mathbf{G}^<$\\
\rule[2mm]{\columnwidth}{1pt} \\
\label{alg:gl}
\end{algorithm}


The key element controlling the efficiency of the HSC method and
its extension is the definition of clusters in the binary tree.
When the binary tree is replaced by a {\it degenerate} (or pathological)
tree where each parent node has only one associated child node,
the HSC method becomes equivalent to the RGF method
(see, for example, \cite[p. 596]{hetmaniuk2013nested}).
For the HSC algorithm and its extension, the multilevel nested dissection
defines the binary tree and the resulting clusters.
The {\it nested dissection} \cite{George1973aa} approach divides
the domain of grid points or atoms into two uncoupled sub-domains
(clusters of grid points/atoms) with a separator, such that the two
disconnected sub-domains are of roughly the same size and the separator
is minimized.
Multilevel nested dissection repeats this division on each sub-domain,
reducing the clusters down to a size small enough (manageable for an
inversion of the corresponding block matrix), or until the clusters
cannot be divided by nested dissection.
In practice, the METIS graph partition library \cite{Karypis:1998:FHQ:305219.305248}
provides the implementation of the multilevel nested dissection.
This well-established partition library allows the HSC extension approach
to handle any arbitrarily-discretized nanoscale electronic devices.

\subsection{Operation count analysis when simulating 3D brick-like devices}
\label{sec:complexity}

Next we discuss the operation counts for the RGF method and for the HSC-extension.
Consider a cuboid device, covered by a three-dimensional orthogonal
mesh with $N_x$, $N_y$, and $N_z$ grid points per direction.
The discretization of the Hamiltonian is obtained via a 7-point stencil.
The self-energy matrix $\boldsymbol{\Sigma}^{r}$ is assumed to be represented
via a similar 7-point stencil (for example, with a crude diagonal approximation
or with a PML-like approximation \cite{nissen2011optimized}).
The resulting matrix $\mathbf{A}$ is of dimension $N_x \times N_z\times N_y$,
where the $y$-direction is the transport direction.

\begin{figure}[hbp]
\centering
\includegraphics[width=0.8\columnwidth]{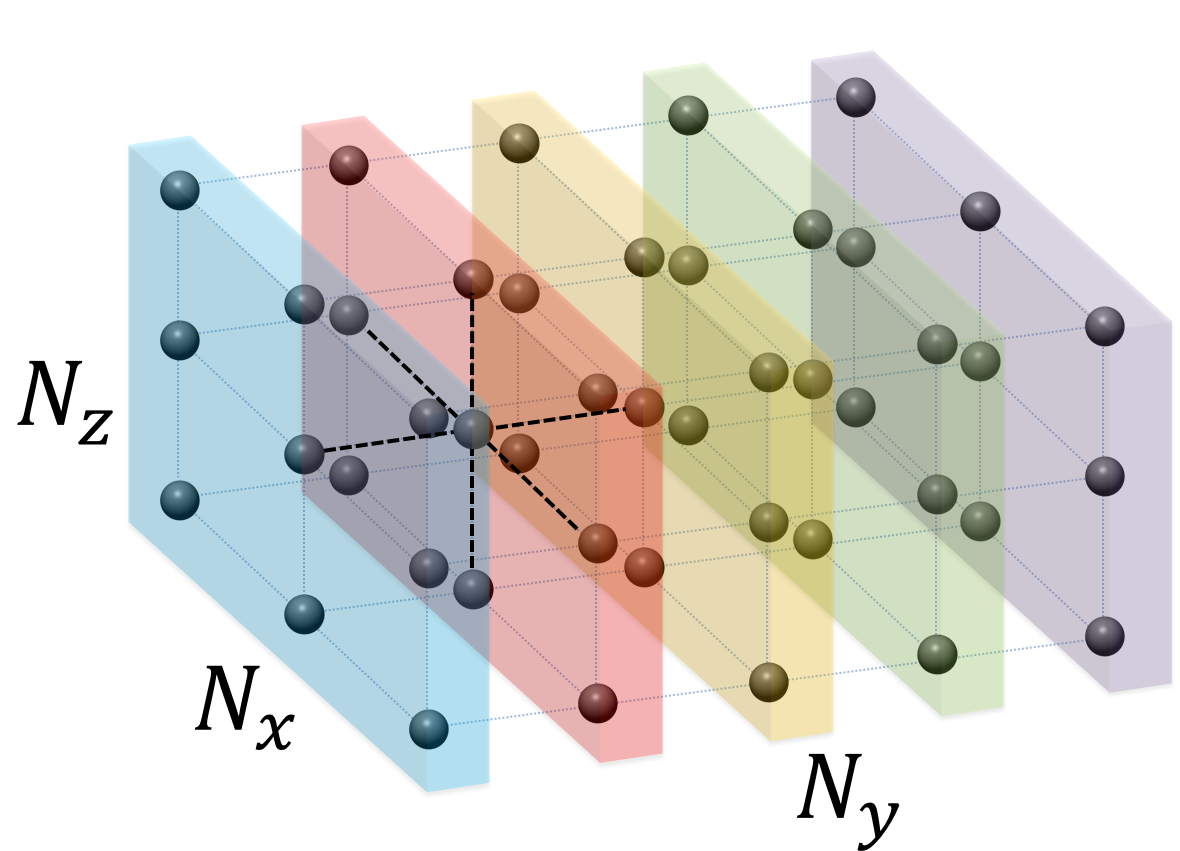}
\caption{A Cartesian 3D mesh with 7-point-stencil discretization of dimension $N_x \times N_z \times N_y$ (the $y$-direction is the transport direction.).
The colored layers along the $y$-direction show the layered-structure organization of grid points for the RGF approach.}
\label{fig:3d_opcount_layers}
\end{figure}

The RGF approach groups the grid points into $N_y$ disjoint layers,
each layer holding $N_x \times N_z$ grid points.
Fig.~\ref{fig:3d_opcount_layers} illustrates these layers for
$N_x = 3$, $N_z = 3$, and $N_y = 5$.
By ordering the grid points one layer at a time, the matrix $\mathbf{A}$
exhibits the block-tridiagonal structure, required by the RGF approach.
The operation count for the RGF method on this cuboid device is
$\mathcal{O}\left( N_x^3 N_z^3 N_y \right)$.

\begin{figure}[hbp]
\centering
\includegraphics[width=\columnwidth]{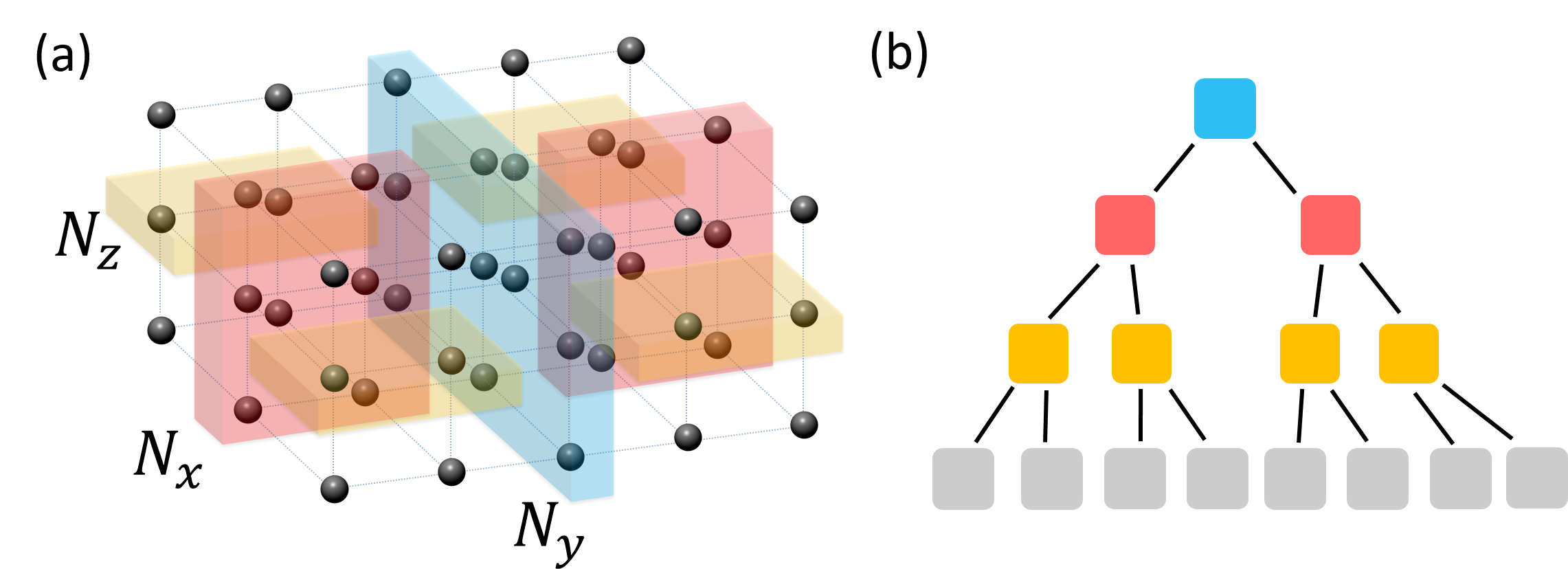}
\caption{(a) The partition of grid points obtained from
the multilevel nested dissection.
The colored clusters show the separators defined at each level.
(b) A binary tree representing the clustering.
Each colored block matches a colored separator.}
\label{fig:3d_opcount_nd}
\end{figure}

The HSC-extension employs a multilevel nested dissection to gather
the grid points into a hierarchy of clusters.
Fig.~\ref{fig:3d_opcount_nd} illustrates the separators obtained
at each level and the eight subdomains.
The resulting binary tree is also depicted in Fig.~\ref{fig:3d_opcount_nd}
with matching colors for the separators.
The operation count for the HSC-extension is derived in the appendix.
For the sake of conciseness, the final value is summarized in Table~\ref{tab:3d_opcount_hsc}.

\begin{table*}[hbpt]
\centering
\begin{tabular}{ccc}
\hline
Configuration  & HSC-extension & RGF \\
\hline
\hline
Cubic mesh ($N_x = N_z = N_y = N$) & $\mathcal{O}(N^6)$ & $\mathcal{O}(N^7)$ \\
Elongated mesh ($N_x = N_z = N \ll N_y$) & $\mathcal{O}(N^5N_y)$ & $\mathcal{O}(N^6N_y)$ \\
Flattened mesh ($N_z\ll  N_x = N_y = N$) & $\mathcal{O}(N_z^3N^3)$ & $\mathcal{O}(N_z^3N^4)$\\
\hline
\end{tabular}
\caption{Operation counts of HSC-extension and RGF for various configurations of cuboid mesh.}
\label{tab:3d_opcount_hsc}
\end{table*}

\begin{Remark}
When $N_z = 1$, the operation counts for RGF and HSC reduce
to their expression for 2D devices (with mesh $N \times N$).
Namely, for RGF, the operation count becomes $\mathcal{O}(N^4)$
and, for HSC, $\mathcal{O} (N^3)$.
\end{Remark}

In practice the self-energy matrix $\boldsymbol{\Sigma}^{r}$
contains dense blo\-cks for the grid points on open boundary
conditions.
The analysis in the appendix and the counts (Table~\ref{tab:3d_opcount_hsc})
do not cover such cases.
The next section will study numerically the operation counts
for practical nanoscale devices with open boundary conditions.

\section{Numerical experiments}
\label{sec:results}

Next we demonstrate and analyze numerically the performance of
HSC-extension approach when evaluating entries of the matrices
$\mathbf{G}^{r}$ and $\mathbf{G}^{<}$.
First a cuboid device is used to illustrate the cost analysis presented
in Section \ref{sec:complexity}.
The impact of dense blocks in $\boldsymbol{\Sigma}^{r}$ when
modeling open boundary conditions is also discussed.
Then three nanoscale devices of practical importance are
considered: a graphene-boron nitride-graphene multilayer system,
a silicon nanowire (SiNW), and a DNA molecule.
The discretizations for these three devices yield matrices with different sparsity pattern and provide different challenges for the HSC-extension.

Both algorithms (RGF and HSC-extension) are implemented as C codes.
All the runtime data corresponds to the total CPU time for the evaluation
of the diagonal and desired off-diagonal entries of $\mathbf{G}^r$ and $\mathbf{G}^<$ at a single energy point.
For reference, timings for the LU-factorization (\texttt{lu} routine in MATLAB 2011b \cite{matlab} calling UMFPACK v5.0.4 \cite{umfpack}) of $\mathbf{A}$ are
also included.
The operation count for the LU-factorization represents the optimal
cost complexity because every solution of a linear system with $\mathbf{A}$
requires, at least, the cost of one LU-factorization.
All numerical experiments are performed with one thread on a machine with Intel i7-2600
3.40GHz CPU and 12GB memory.

\subsection{Cuboid nanoscale device}

As described in section \ref{sec:complexity}, a cuboid nanoscale device
is considered where the Hamiltonian is constructed by effective-mass
approximation and with 7-point stencil finite difference.
A three-dimensional orthogonal mesh is used with $N_x$, $N_y$, and
$N_z$ grid points per direction.
Two distinct treatments for the self-energy matrices are studied:
a diagonal approximation, referred to as {\it SPARSE}, and a {\it DENSE}
approximation for modeling open boundary conditions.

\subsubsection{Results with {\it SPARSE} self-energy matrices}

\begin{figure}[hbp]
\centering
\includegraphics[width=0.7\columnwidth]{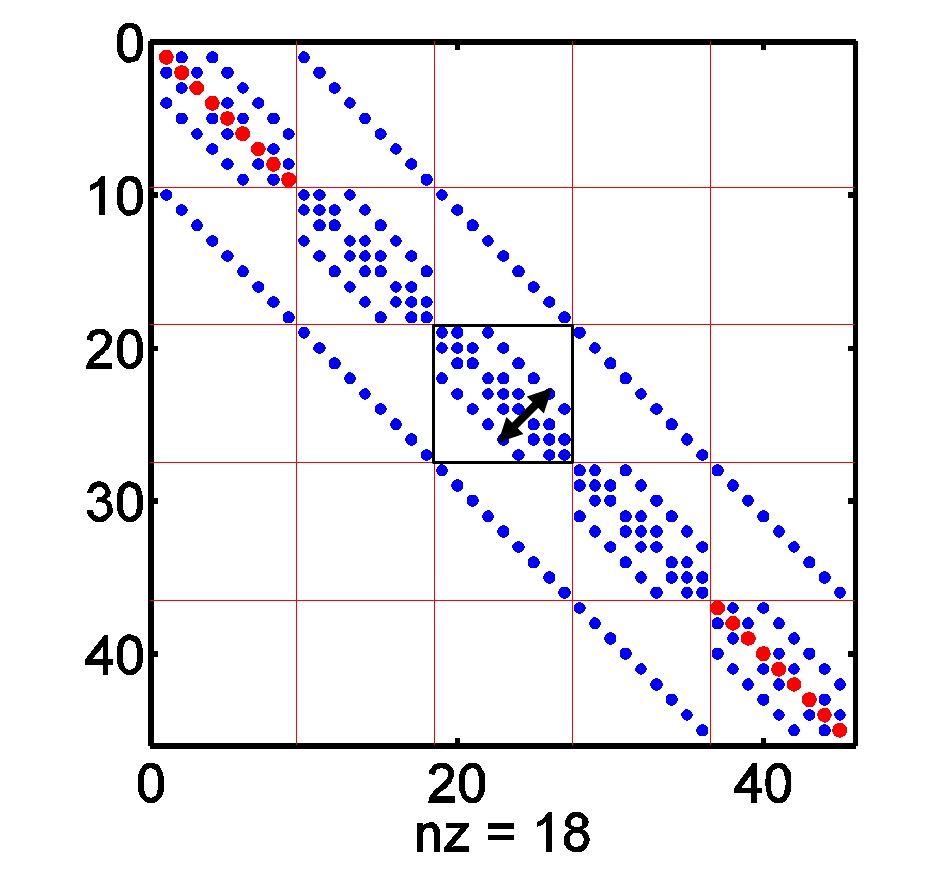}
\caption{Non-zero pattern of $\mathbf{A}$ for a 3D cuboid system with
$N_x= 3$, $N_z = 3$ and $N_y = 5$.
Entries for the diagonal self-energy approximation are marked
in red.
The matrix exhibits a block-tridiagonal structure, where each block
is of dimension $N_xN_z\times N_xN_z$.
The arrow highlights the diagonal width in each block controlled by
the ratio $N_x/N_z$. In all the matrix pattern graphs,
\texttt{nz} specifies the number of non-zero entries.}
\label{fig:cuboid_sparseSigma_matrix}
\end{figure}

Here diagonal self-energy matrices are considered.
The matrix $\mathbf{A}$ has the same sparsity as the Hamiltonian
matrix $\mathbf{H}$.
Fig.~\ref{fig:cuboid_sparseSigma_matrix} illustrates the pattern
of non-zero entries in the matrix $\mathbf{A}$, when the grid
points are ordered one layer at a time (as in Fig.~\ref{fig:3d_opcount_layers}).

\begin{figure}[hbp]
\centering
\includegraphics[width=0.7\columnwidth]{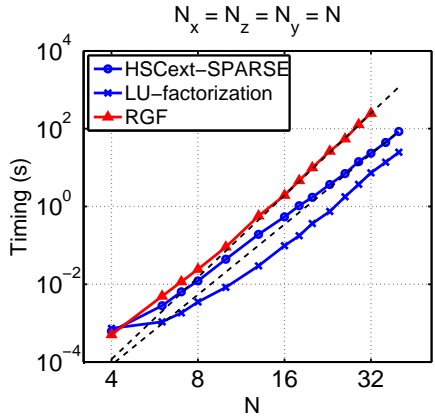}
\caption{CPU timing for cubic system versus the dimension $N_x$. For all plots in result section, the timing includes the $\mathbf{G}^r$ and $\mathbf{G}^<$ calculation at one energy point. The runtime of RGF, HSC-extension and LU-factorization for $\mathbf{A}$ with {\it SPARSE} self-energy are presented. For comparison, we also plot black dashed curves, reflecting the theoretically asymptotic slopes for HSC-extension: $\mathcal{O}(N^6)$, and for RGF: $\mathcal{O}(N^7)$.}
\label{fig:cuboid_time_cube}
\end{figure}

First, when $N_x = N_y = N_z = N$, the CPU times for evaluating $\mathbf{G}^{r}$ and $\mathbf{G}^{<}$
at one energy point are plotted as a function of $N$
in Fig.~\ref{fig:cuboid_time_cube}.
The slopes are consistent with the analysis of section \ref{sec:complexity},
namely $\mathcal{O}(N^7)$ for RGF and $\mathcal{O}(N^6)$ for HSC-extension.
The LU-factorization of $\mathbf{A}$ exhibits also a complexity
$\mathcal{O}(N^6)$.
When $N = 32$, the HSC-extension exhibits a speed-up of
10 times.
Note that, on this 12GB machine, RGF can solve problems
only up to $N = 32$ (the resulting matrix $\mathbf{A}$ is of
dimension 32,768), while the HSC-extension can solve these problems
up to $N = 40$ (dimension of matrix $\mathbf{A}$ is 64,000).

\begin{figure}[hbp]
\centering
\includegraphics[width=\columnwidth]{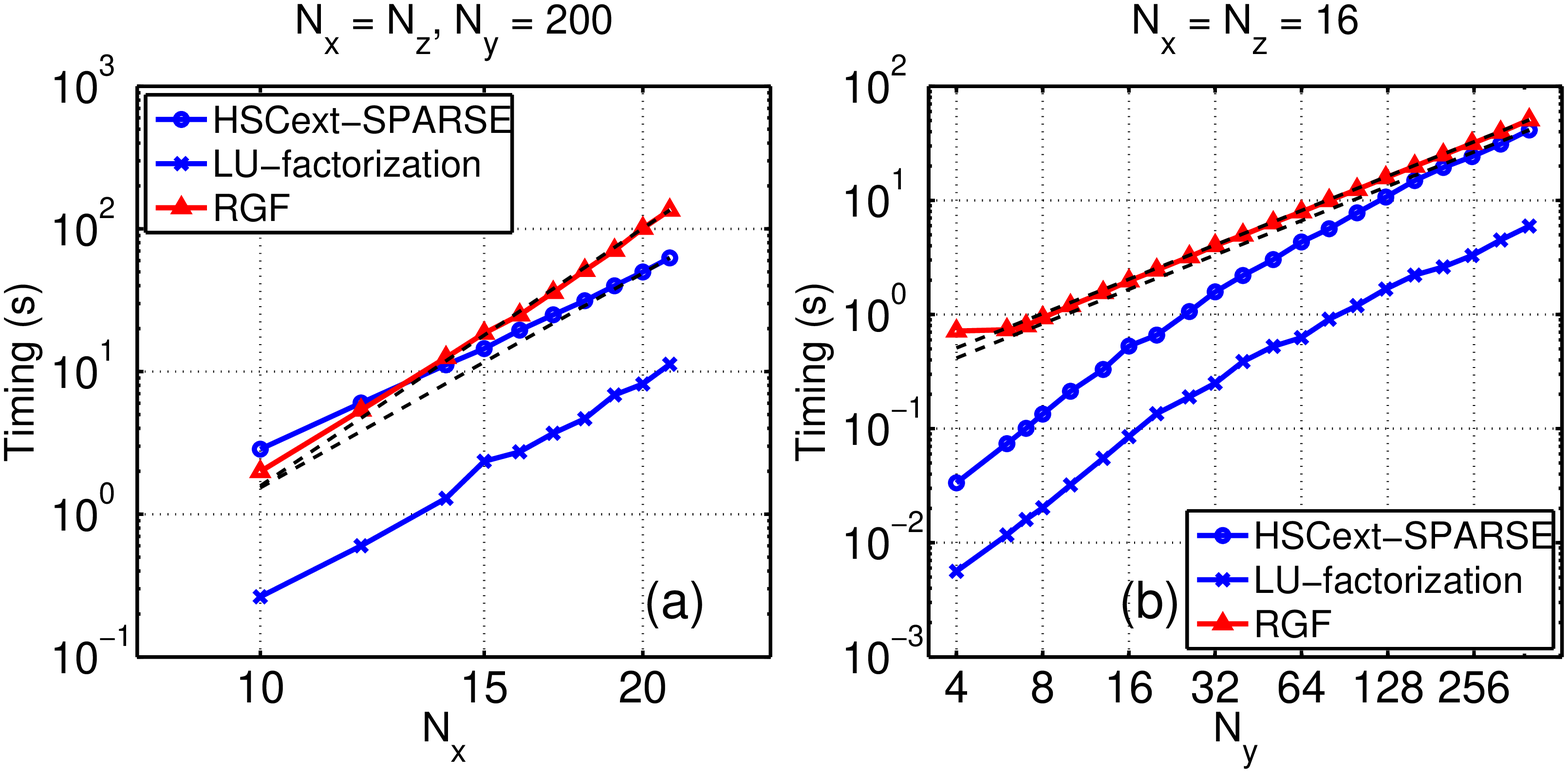}
\caption{(a) CPU timing for elongated mesh versus $N_x$ with fixed $N_x = N_z =N \ll N_y $, $N_y=200$. (b) CPU timing for elongated mesh versus $N_y$ with fixed $N_x = N_z =16$.
The theoretically asymptotic slopes (black dashed curves) for HSC-extension correspond to Table~\ref{tab:3d_opcount_hsc}, $\mathcal{O}(N^5 N_y)$, and for RGF
to $\mathcal{O}(N^6 N_y)$.}
\label{fig:cuboid_elong}
\end{figure}

Next the case of an elongated device is considered, {\it i.e.}
$N_x = N_z = N \ll N_y$.
Fig.~\ref{fig:cuboid_elong} illustrates timings for different
elongated devices with square cross-section ($N_x = N_z = N$).
The asymptotic slopes (black dashed curves) match the analysis,
namely $\mathcal{O}(N^5 N_y)$ for the HSC-extension
and $\mathcal{O}(N^6 N_y)$ for RGF.
Here again the HSC-extension and the LU-factori\-za\-tion have
numerically the same complexity.
When $N_y$ is fixed at 200, the CPU time of HSC-extension is initially
higher than the one for RGF at small cross-sections and becomes
smaller than the one for RGF when $N_x = N_z \geq 12$, eventually reaching
a speed-up of 2 for the largest structure studied.

\begin{figure}[hbp]
\centering
\includegraphics[width=0.7\columnwidth]{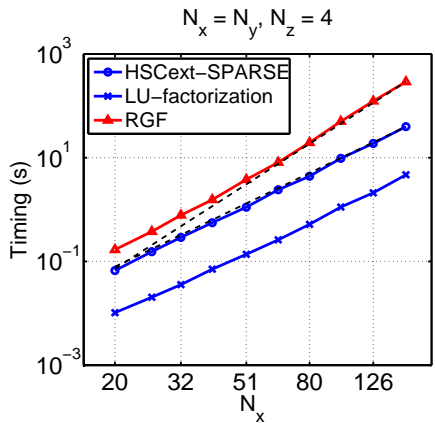}
\caption{CPU timing for flattened mesh versus $N_x$ with $N_x = N_y = N \gg N_z$, $N_z = 4$. The theoretically asymptotic slope (black dashed curves) for flattened mesh is $\mathcal{O}(N^3)$ for HSC-extension, and $\mathcal{O}(N^4)$ for RGF.}
\label{fig:cuboid_flat}
\end{figure}

Finally, for the case of flattened devices, the CPU results are shown
in Fig.~\ref{fig:cuboid_flat}.
When $N_z = 4$ and $N_x = N_y = N\gg N_z$, the costs of $\mathcal{O}(N^3)$
for HSC-extension and $\mathcal{O}(N^4)$ for RGF are observed.
These asymptotic behaviors are consistent with the analysis
in section \ref{sec:complexity} and with the conclusions
for 2D devices with $N_x \times N_y$ grid points
\cite{hetmaniuk2013nested}.

Our numerical experiments in Fig.~\ref{fig:cuboid_sparseSigma_matrix}-\ref{fig:cuboid_flat}
illustrate the asymptotic operation count of HSC-extension as a function of
system dimensions for various cuboidal shapes.
In all three cases, the HSC-extension and the LU-facto\-riza\-tion have
identical asymptotic operation counts.
These numerical experiments strongly suggest that the HSC-extension
reaches the ideal complexity for 3D nanoscale devices.

\subsubsection{Effect of {\it DENSE} self-energy matrices}

\begin{figure}[hbp]
\centering
\includegraphics[width=0.7\columnwidth]{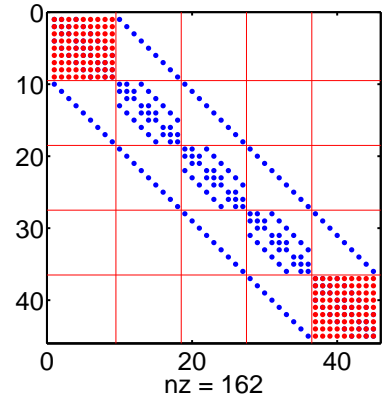}
\caption{Non-zero pattern of $\mathbf{A}$ for a 3D cuboid system with
$N_x= 3$, $N_z = 3$ and $N_y = 5$.
Entries for the diagonal self-energy approximation are marked
in red.
The matrix exhibits a block-tridiagonal structure, where each block
is of dimension $N_xN_z\times N_xN_z$.}
\label{fig:cuboid_denseSigma_matrix}
\end{figure}

Next dense self-energy matrices are considered to model open boundary
conditions.
Fig.~\ref{fig:cuboid_denseSigma_matrix} illustrates the pattern
of non-zero entries in the matrix $\mathbf{A}$, when the grid
points are ordered one layer at a time
(as in Fig.~\ref{fig:3d_opcount_layers}).

RGF does not exploit the sparsity present in most diagonal blocks of matrix $\mathbf{A}$.
So using a dense self-energy matrix does not impact the performance of RGF.
On the other hand, the HSC-extension aims to exploit as much as
possible the sparsity of $\mathbf{A}$.
So it is important to study the impact of a dense self-energy matrix
on the HSC-extension.
The analysis in section \ref{sec:complexity} does not handle this situation.

\begin{figure}[hbp]
\centering
\includegraphics[width=0.7\columnwidth]{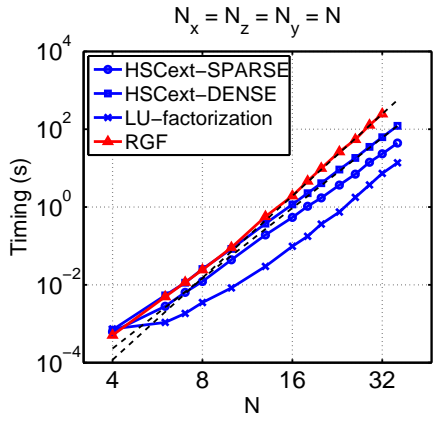}
\caption{CPU timing for cubic system $N_x = N_y = N_z = N$ with
both {\it DENSE} and {\it SPARSE} self-energies.
The black dashed curve shows the asymptotic rates,
namely $\mathcal{O}(N^6)$ for HSC-extension and $\mathcal{O}(N^7)$
for RGF.}
\label{fig:cuboid_dense}
\end{figure}

First consider the case where $N_x = N_y = N_z = N$.
Fig.~\ref{fig:cuboid_dense} plots the CPU timings as a function
of $N$.
Here the RGF calculation stops at $N = 32$ due to memory limitation,
while the HSC-extension can solve problems up to $N = 36$ with
dense self-energy matrices.
The pre\-sence of a dense self-energy matrix yields larger CPU times
for the HSC-extension but the asymptotic operation count is
not modified.

\begin{figure}[hbp]
\centering
\includegraphics[width=0.7\columnwidth]{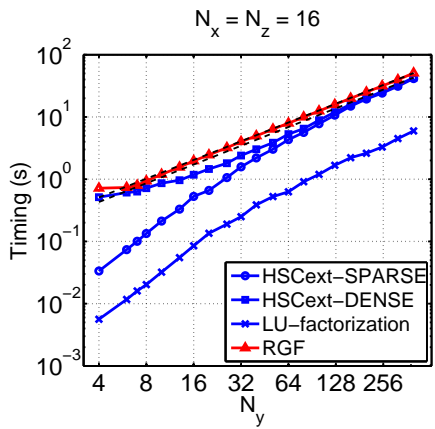}
\caption{CPU timing for different $N_y$ and fixed $N_x = N_z = 16$.
The black dashed curve shows the asymptotic rate $\mathcal{O}(N_y)$.}
\label{fig:cuboid_dense_elongated}
\end{figure}

Fig.~\ref{fig:cuboid_dense_elongated} plots the CPU timings
when $N_y$ is modified and $N_x$ and $N_z$ remain constant.
Timings for RGF, for HSC-exten\-sion with sparse self-energy matrix,
for HSC-extension with dense self-energy matrix, and for
the LU-factorization of $\mathbf{A}$ are reported.
Note that in this numerical experiment, the layered-structure decomposition employed in RGF is kept along $y$-direction (even when $N_y < N_x$).
When $N_y$ is comparable to $N_x$ and $N_z$, the speed-up for HSC-extension
over RGF is reduced when a dense self-energy matrix is considered.
As $N_y$ gets larger, the timings of HSC-extension with the two forms
of self-energy are closer and, asymptotically, approaching
the complexity $\mathcal{O}(N_y)$.

\subsection{Graphene - boron nitride - graphene multilayer system}

\begin{figure}[hbp]
\centering
\includegraphics[width=\columnwidth]{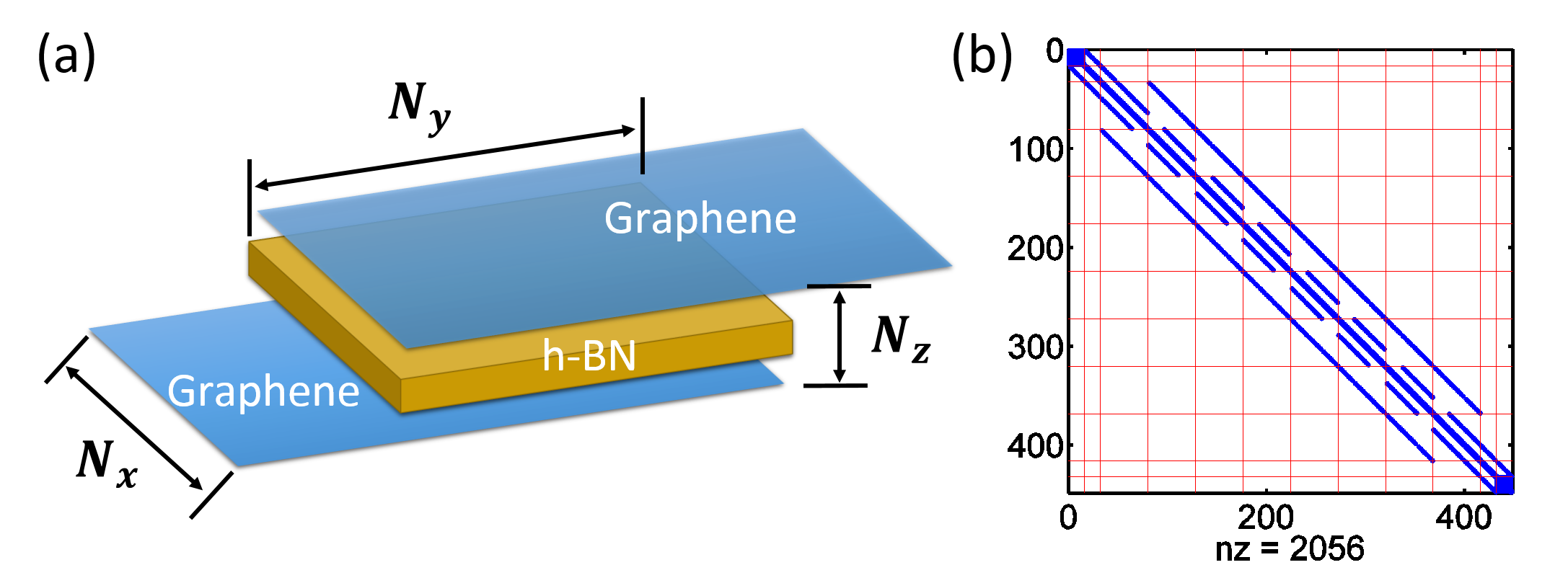}
\caption{(a) Schematic view of graphene-hBN-graphene multilayer heterostructure. Two graphene layers are semi-infinitely long used as contacts. (b) The non-zero pattern of the $\mathbf{A}$ matrix with $N_x = 16$, $N_y = 8$ and $N_z = 3$.}
\label{fig:gbng}
\end{figure}

Graphene, stacked with boron nitride insulating material, is a promising material to build next generation transistors because of its extraordinary thermal and electronic properties \cite{gastro2009}.
Here a multilayer heterostructure is considered, as shown in
Fig.~\ref{fig:gbng}.

The device consists of two semi-infinitely long monolayer armchair-edged
graphene nanoribbon (AGNR) electrodes sandwiching an ultra-thin hexagonal
boron nitride (h\-BN) multilayer film, yielding a vertical
tunneling heterostructure with hBN acting as a potential
barrier \cite{britnell2013resonant}.
The hBN film is a few atomic layers thick and the central graphene-hBN-graphene
(G-BN-G) overlapping hetero\-struc\-ture/multi\-layer region is stacked in
AB-order (Bernal stacking).
For this problem, the number of 2D vertical layers is denoted by $N_z$
in units of atomic layers.
The system width is $N_x$ and the length of the multilayer stacking region
is $N_y$, also in units of atomic layers.
The semi-infinitely long AGNR monolayer electrodes at the top and bottom layers
are treated as open boundary conditions, their effect is folded into dense
self-energy blocks (extreme blocks) of dimension $N_x \times N_x$.

The system Hamiltonian is constructed using the nea\-rest neighbor tight
binding approximation with parameters from \cite{zhao2015negative}.
Only the low energy $p_z$ orbitals are considered here;
thus the Hamiltonian has the same dimension as the total number of atoms
simulated.
The geometric lattice complexity of the multilayer system yields
an average 5-point stencil Hamiltonian sparsity
(multiple hexagonal-meshed layers stacked in AB order).
The complexity of RGF remains
$\mathcal{O}\left(N_{x}^{3}N_z^3N_{y} \right)$
because the sparsity inside each block is not exploited.
The complexity of HSC-extension is studied numerically.

\begin{figure}[hbp]
\centering
\includegraphics[width=0.7\columnwidth]{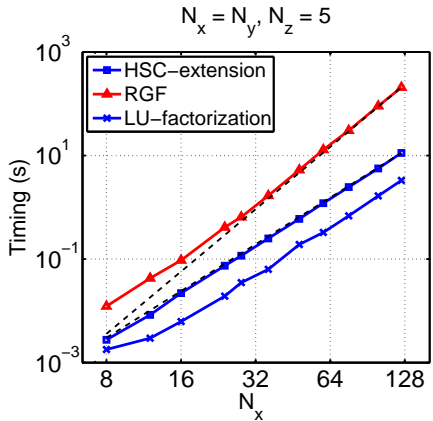}
\caption{CPU timing for G-BN-G system as a function of $N_x = N_y$ and fixed $N_z=5$.
Dashed curves illustrate asymptotic rates, namely
$\mathcal{O}(N_x^3)$ for HSC-extension and $\mathcal{O}(N_x^4)$
for RGF.}
\label{fig:gbng_square}
\end{figure}

Fig.~\ref{fig:gbng_square} plots the CPU timings
when $N_x = N_y$ and $N_z = 5$.
The HSC-extension and the LU-factorization of $\mathbf{A}$ have
the same asymptote, indicating an operation count
$\mathcal{O}(N_x^3)$.

\begin{figure*}
\centering
\includegraphics[width=0.8\textwidth]{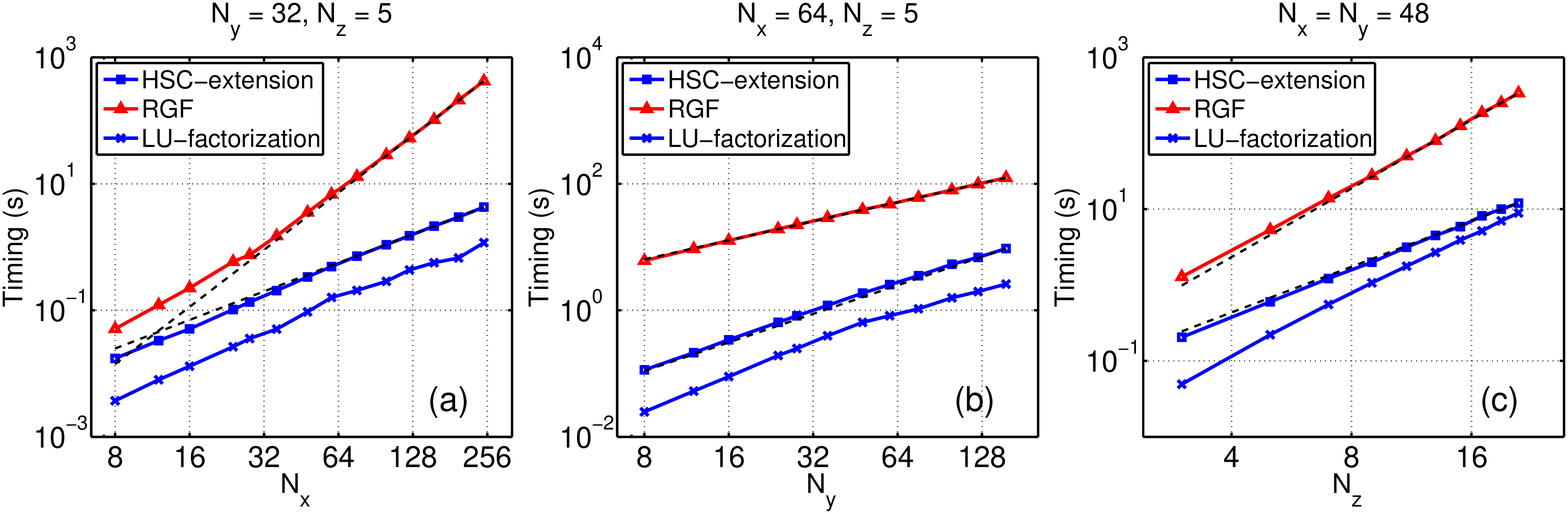}
\caption{(a) CPU timings for G-BN-G system as a function of $N_x$ and fixed $N_y=32$, $N_z=5$.
(b) CPU timings for different $N_y$ and fixed $N_x=64$, $N_z=5$.
(c) CPU timings as a function of $N_z$ and fixed $N_x = N_y = 48$.
The dashed curves indicates the asymptotic operation counts.
For HSC-extension, they are $\mathcal{O}(N_x^{1.5})$ for (a),
$\mathcal{O}(N_y^{1.5})$ for (b), and $\mathcal{O}(N_z^{2})$ for (c).
The operation counts for RGF are as follows: $\mathcal{O}(N_x^3)$ for (a),
$\mathcal{O}(N_y)$ for (b), and $\mathcal{O}(N_z^{3})$ for (c).
}
\label{fig:gbng_xyz}
\end{figure*}

Fig.~\ref{fig:gbng_xyz} plots the CPU timings for different
configurations of $N_x$, $N_y$, and $N_z$.
The experiments illustrate that HSC-extension still exhibits
a complexity similar to the LU-factori\-za\-tion of $\mathbf{A}$.
For the largest devices simulated in Fig.~\ref{fig:gbng_xyz}(a),
$N_x = 256$, the HSC-extension method offers a speed-up of 3 orders
of magnitude over RGF.

The numerical experiments indicate that the asymptotic cost of HSC-extension
is $\mathcal{O}\left( N_{x}^{1.5}N_z^2N_{y}^{1.5} \right)$.
This cost of HSC-extension can be compared to the cost
(Table~\ref{tab:3d_opcount_hsc}) for the flattened device.
The term $\mathcal{O}(N_x^{1.5} N_y^{1.5})$ matches well with
$\mathcal{O}(N^3)$ by assuming $N_x = N_y = N$.
The term $\mathcal{O}(N_z^2)$ is due to the Bernal stacking order
for the multilayer structure in the $z$-direction.
\begin{figure}[hbp]
\centering
\includegraphics[width=\columnwidth]{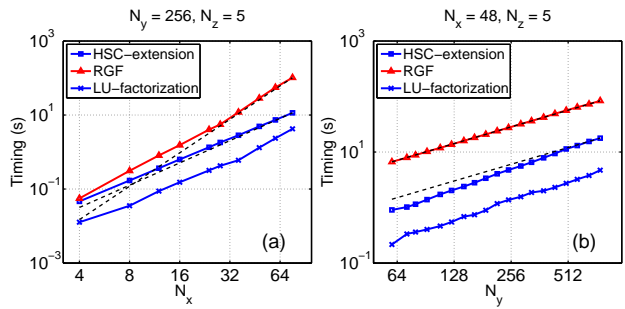}
\caption{(a) CPU timing for G-BN-G system with different $N_x$ and fixed $N_y=256$, $N_z=5$. (b) CPU timing for different $N_y$ and fixed $N_x=48$, $N_z=5$.
The dashed curves indicates the asymptotic operation counts.
For HSC-extension, they are $\mathcal{O}(N_x^{2})$ for (a) and
$\mathcal{O}(N_y)$ for (b).
The operation counts for RGF are as follows: $\mathcal{O}(N_x^{3})$ for (a)
and $\mathcal{O}(N_y)$ for (b).
}
\label{fig:gbng_elong}
\end{figure}

Finally, Fig.~\ref{fig:gbng_elong} illustrates CPU timings
for an elongated device, {\it i.e.} $N_x \ll N_y$.
The timings for the HSC-extension behave like $\mathcal{O}(N_x^2N_y)$,
demonstrating a lower order of complexity over the RGF method.

As a summary, the runtime cost of HSC-extension for the G-BN-G multilayer
structure is
\begin{equation}
T =
  \begin{cases}
   \mathcal{O}(N_x^{1.5} N_y^{1.5} N_z^2) & \quad \text{when}~ N_z \ll N_x \simeq N_y
   \\
   \mathcal{O}(N_x^{2} N_y N_z^2)  & \quad \text{when}~ N_z \ll N_x \ll N_y
  \end{cases}
\label{eq:gbngcost}
\end{equation}
while the runtime cost of RGF behaves like $\mathcal{O}(N_x^3 N_z^3 N_y)$.

\subsection{Silicon nanowire structure}

Silicon nanowire devices have shown promises to become key components
in the next generation computer chips \cite{cui2001functional}.
Solving efficiently the NEGF equations for such devices is therefore important.

\begin{figure}[hbp]
\centering
\includegraphics[width=\columnwidth]{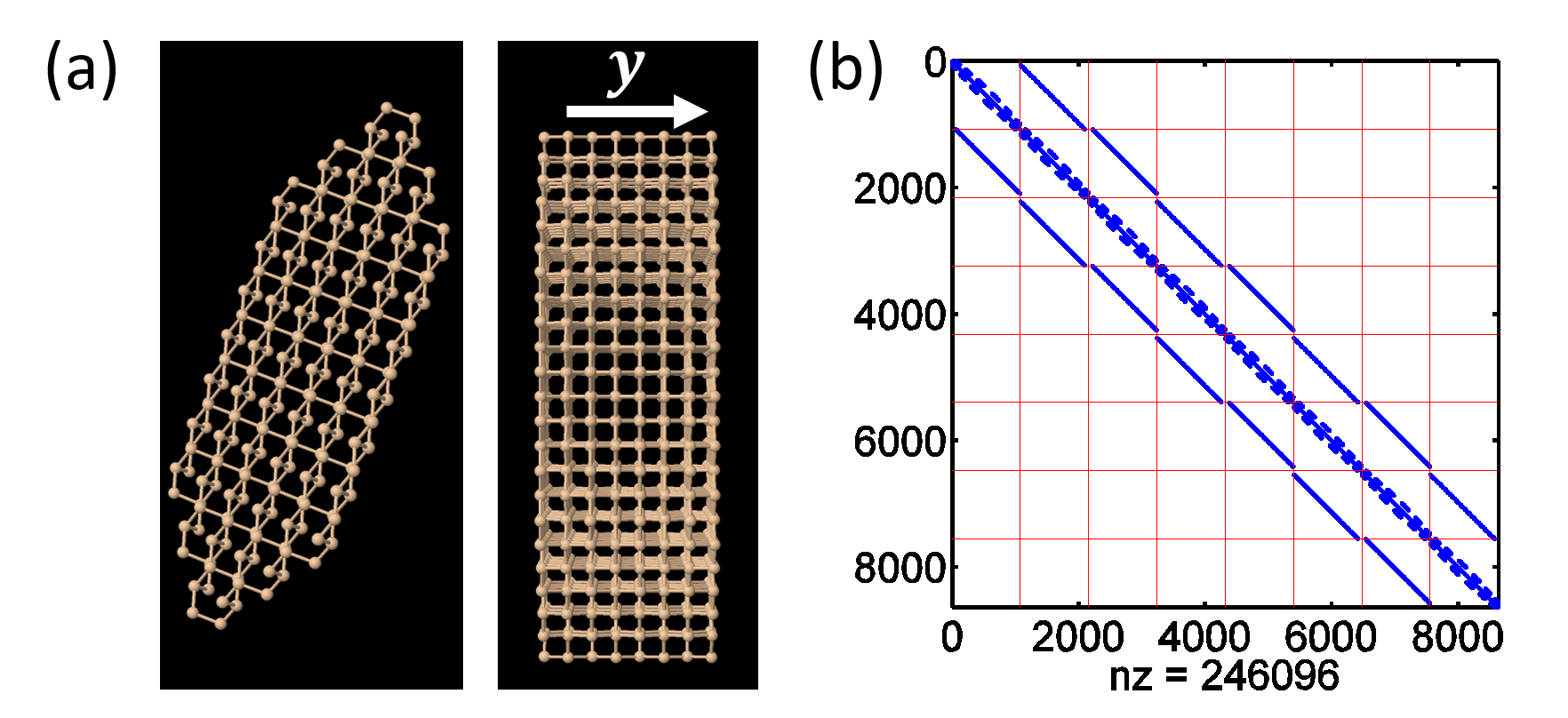}
\caption{(a) Atomic view of a silicon nanowire example with 4 unit cells.
Each unit cell has two atomic layers and hexagonal cross-section shape,
with each atomic layer containing 108 Si atoms. Cross-section is along $x-z$
plane and transport direction is along $y$ direction.
This example corresponds to $N_{cs} = 108$ and $N_y = 8$.
(b) The non-zero pattern of the $\mathbf{A}$ matrix with $N_{cs} = 108$ and $N_y = 8$.}
\label{fig:nanowiregolam}
\end{figure}

In order to investigate the scaling of computational runtime as a function of SiNW lateral dimensions, specifically the number of atoms in each layer and the number of unit cells, we consider a SiNW device depicted in Fig.~\ref{fig:nanowiregolam}(a).
The number of atomic layers in $y$-direction is denoted as $N_y$ and the number of silicon atoms within each atomic layer (cross-section) is denoted as $N_{cs}$.
So a $N_{cs} \times N_y$ SiNW structure contains an array of $N_y/2$ unit
cells with $2N_{cs}$ atoms per unit cell.
Next the $sp^3d^5s^*$ tight-binding formalism \cite{luisier2006atomistic}
is used to discretize the system.
Each silicon atom is represented by a $10\times 10$ diagonal block, thereby interconnecting with up to 40 orbitals of the nearest-neighbor silicon atoms.
The resulting Hamiltonian matrix exhibits a stencil involving more than
40 points, which results in a particular computational challenge.
Dense self-energy matrices of dimension $10N_{cs} \times 10N_{cs}$
are employed.
Fig.~\ref{fig:nanowiregolam}(b) illustrates the sparsity of $\mathbf{A}$
when $N_{cs} = 108$ and $N_y = 8$.

In this section, we consider nanowire whose length $L$ is proportional to the diameter $D$ of the cross-section, namely $L = \alpha D$.
When ordering the atoms one layer at a time, the Hamiltonian matrix,
as well as matrix $\mathbf{A}$, has a block-tridiagonal structure,
where each block is of dimension $40N_{cs} \times 40N_{cs}$.
The operation count for RGF becomes $\mathcal{O}\left( N_{cs}^3 N_y \right)$.
Since $N_y = \mathcal{O}(L)$, $N_{cs} = \mathcal{O}( D^2 )$, and $L = \alpha D$, the operation count for RGF is $\mathcal{O}(N_y^7)$.
The operation count for HSC-extension will be studied numerically.

\begin{figure*}
\centering
\includegraphics[width=0.8\textwidth]{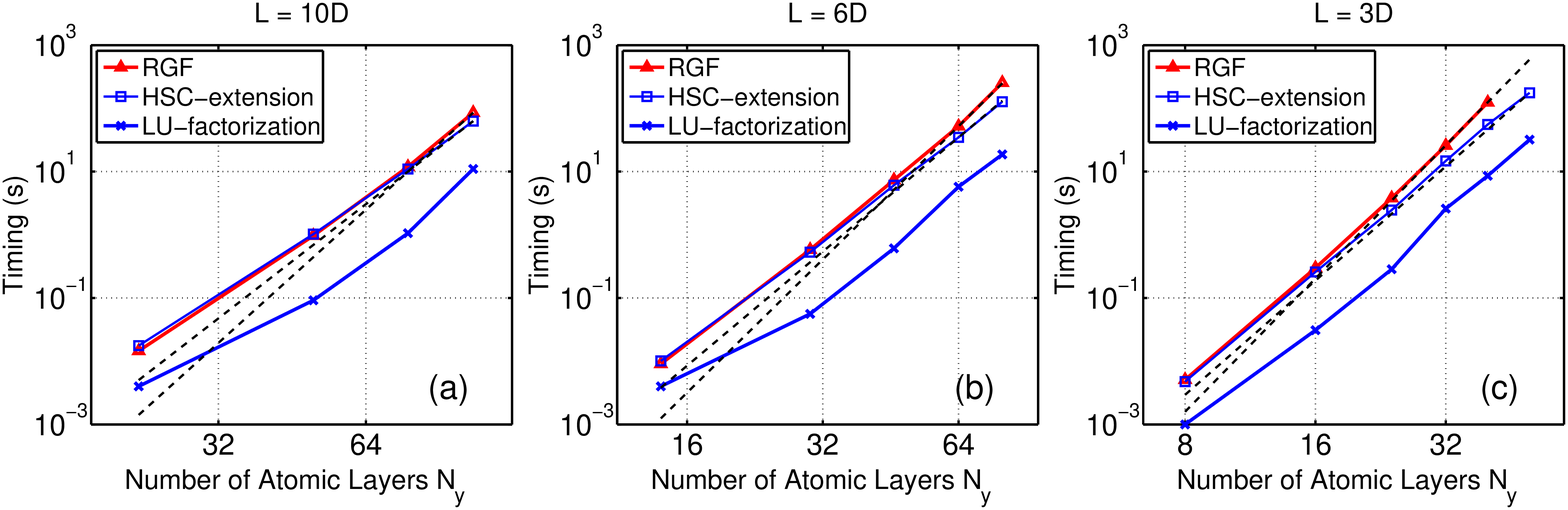}
\caption{CPU timing for SiNW system with (a) $L = 10D$ with largest $L=20$nm,
(b) $L = 6D$  with largest $L=15$nm and (c) $L = 3D$ with largest $L = 9.3$nm.
Note that $N_y \propto L$ and $N_{cs} \propto D^2$.
The dashed curves represent asymptotes:
$\mathcal{O}(N_{y}^{6})$ for HSC-extension and $\mathcal{O}(N_y^7)$ for RGF.}
\label{fig:nanowire}
\end{figure*}
Fig.~\ref{fig:nanowire} plots CPU timings as a function of $N_y$ for structures shaped by
$L = 10D$, $L = 6D$ and $L = 3D$.
Our numerical experiments exhibit an asymptotic cost of $\mathcal{O}(N_{y}^{7})$
for RGF and $\mathcal{O}(N_{y}^{6})$ for HSC-extension.
We would like to emphasize that the complexity $\mathcal{O}(N_{y}^{6})$ is valid for HSC-extension
as long as $L = \alpha D$, independent of the value $\alpha$.
The HSC-extension has the same asymptotic behavior as the LU-factori\-za\-tion
of $\mathbf{A}$.

In practice, analysts may consider nanowires of 20nm length.
Table~\ref{tab:sinw} lists CPU timings when the length is 20nm.
Because these simulations did not fit in the 12GB RAM of our machine, the CPU timings were extrapolated from the asymptotes in Fig.~\ref{fig:nanowire}.
\begin{table}[hbpt]
\centering
\begin{tabular}{cccc}
\hline
Shapes  &  HSC-extension (s) & RGF (s) & speed-up\\
\hline
\hline
$L = 10D$  &  62.9 &  85.3 & 1.4 \\
$L = 6D$   &  1,064  & 3,030 &  2.8 \\
$L = 3D$ &  19,920 &  147,706 & 7.4 \\
\hline
\end{tabular}
\caption{Extrapolated CPU timings of transmission calculation for SiNW devices at one energy point
with $L=20$nm for various shapes.\label{tab:sinw}}
\end{table}
This extrapolation indicates that the speedup of HSC-extension over RGF
improves as $\alpha$ decreases. The reduction of $\alpha$ enlarges
the cross-section $N_{cs}$ while $L$ is fixed, yielding a higher speedup of HSC-extension,
which is consistent with the other experiments.

\subsection{DNA molecule}

Finally we test the algorithm for DNA-based structure, which represents a complex organic system. It has been shown that DNA is one of the promising candidates in the molecule devices \cite{Gohler18022011}.
The study of electronic structures can be used to develop new sequencing techniques \cite{postma2010rapid}, acting as DNA fingerprints.
Another application is disease detection \cite{tsutsui2011electrical}. Many diseases are linked with the mutation in DNA bases, resulting in different electronic properties which can be used to distinguish the mutated DNA.
In our numerical experiments, the DNA molecule is described by density functional theory (DFT) method.
Although the number of atoms contained in single DNA molecule is not huge, the decomposed Hamiltonian matrices are relatively dense, thus impeding an effective decomposition using the multilevel nested dissection.

\begin{figure}[hbp]
\centering
\includegraphics[width=\columnwidth]{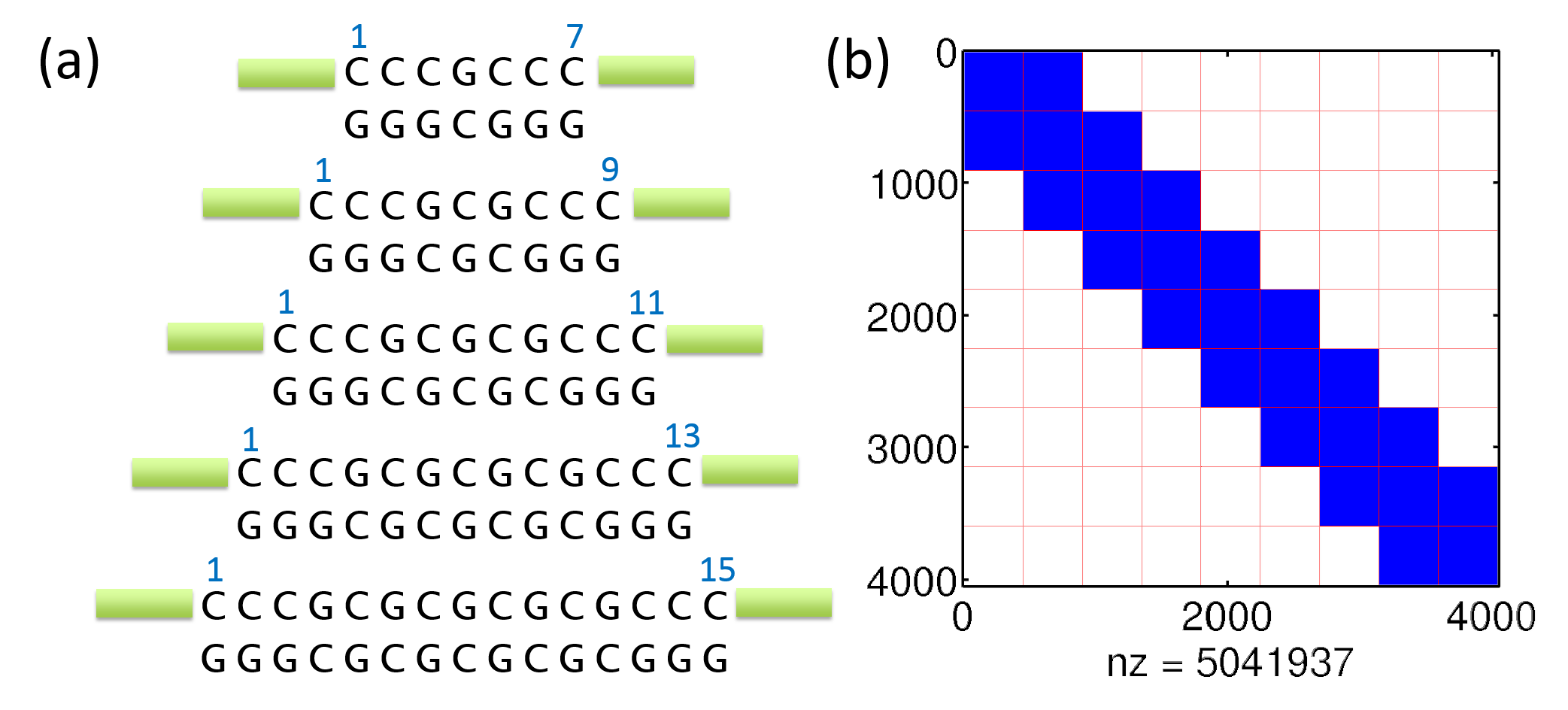}
\caption{(a) Sketch of the simulated DNA sequence with 7, 9, 11, 13 and 15 base pairs respectively. Cytosine (C) and guanine (G) are two types of bases in DNA. The left/right contacts are connected to the bases on one strand. (b) The corresponding non-zero pattern of the $\mathbf{A}$ matrix for the 9 base pairs DNA. All tri-diagonal blocks are fully dense.}
\label{fig:dna}
\end{figure}

The DNA molecule in our simulation is a double-helix structure containing
7 -- 15 base pairs in each strand sketched in Fig.~\ref{fig:dna}(a).
The Hamiltonian matrices are generated by DFT package GAUSSIAN 09 \cite{g09} at HG/6-31G ($d$, $p$) level \cite{qi2013unified}.
The number of orbitals (matrix dimension) for each base is about 250.
For example, for a 9-mer DNA molecule, the Hamiltonian is of dimension $4500\times 4500$ as shown in Fig.~\ref{fig:dna}(b).
The Hamiltonian can be decomposed by treating each base pair as one layer, yielding a block tri-diagonal shape with 9 layers.
Different from the structures studied above, the diagonal and nearest neighbor off-diagonal blocks in the Hamiltonian are fully dense.


The CPU timing results for various DNA molecules are summarized
in Table~\ref{tab:dna}.
\begin{table}[hbpt]
\centering
\begin{tabular}{cccccc}
\hline
Number of Base Pairs & 7 &  9 & 11 & 13 & 15\\
\hline
\hline
HSC-extension (s) & 5.0 & 8.7  & 11.3 & 12.5 & 14.6\\
RGF (s) & 5.5 & 7.8 & 9.7 & 11.6 & 13.7\\
\hline
\end{tabular}
\caption{CPU timings of transmission calculation for DNA molecules
at one energy point.\label{tab:dna}}
\end{table}
For these configurations, the HSC-extension seems to be less efficient
than RGF.

\begin{figure}[hbp]
\centering
\includegraphics[width=\columnwidth]{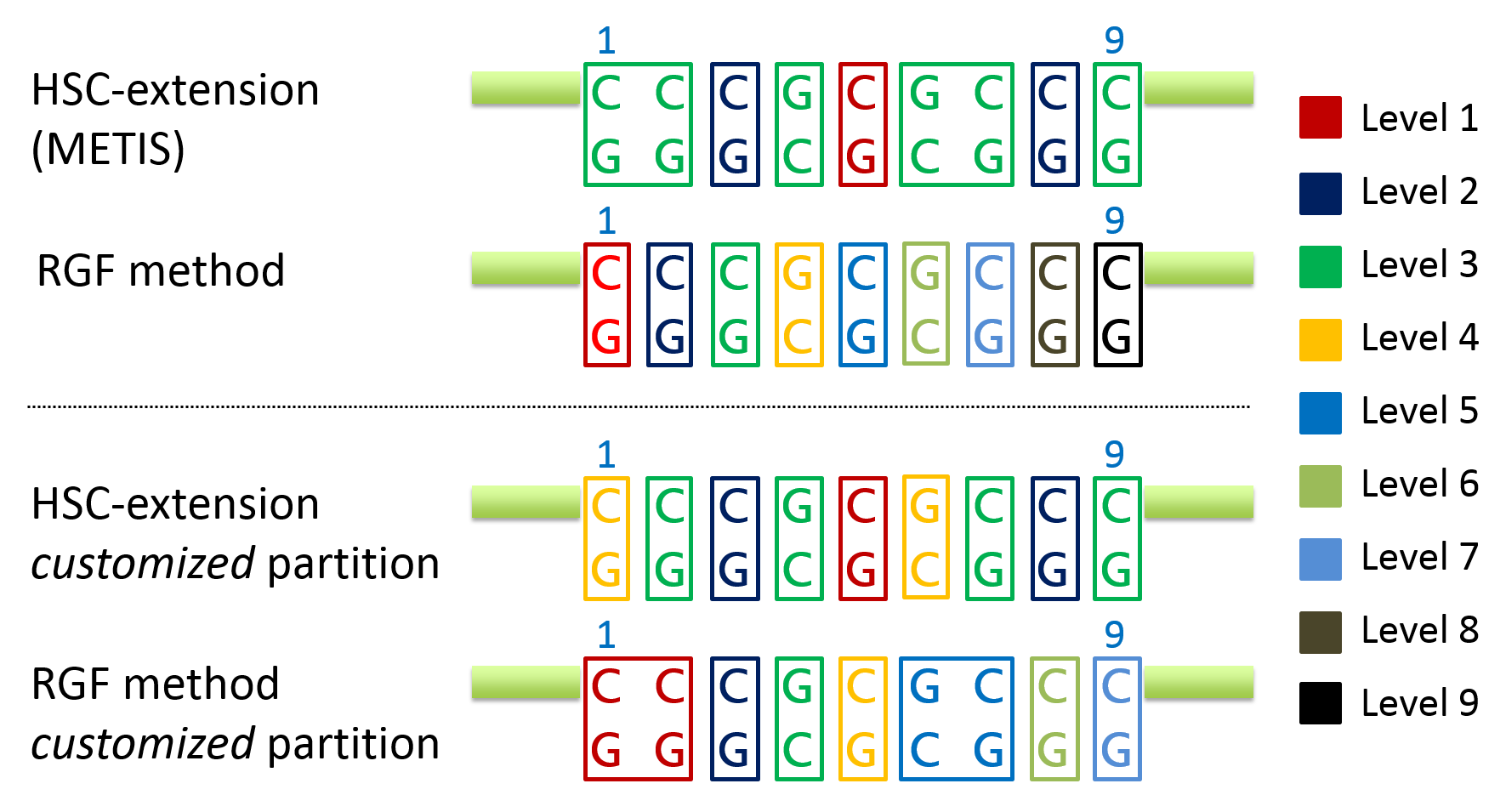}
\caption{Cluster definitions for HSC-extension, for RGF, and for two customizations.}
\label{fig:dna9}
\end{figure}

To better understand these CPU times, it is important to look
at the clusters defined by the multilevel nested dissection.
Fig.~\ref{fig:dna9} illustrates the clusters used for the HSC-extension
when the multilevel nested dissection is applied {\it blindly}
to the graph of $\mathbf{A}$, and the layers used for RGF.
RGF employs layers with one base pair and the resulting blocks
are of dimension $500 \times 500$.
The multilevel nested dissection works at the level of base pairs
because one base pair corresponds to one fully-dense diagonal block
in the matrix $\mathbf{A}$.
The resulting partition introduces clusters with one base pair
except for two bottom level (level 3) clusters that can not be partitioned
with nested dissection.
These two clusters result in two block matrices of dimension
$1000 \times 1000$.
The time discrepancy arises from these two distinct choices of
row gathering.

To further illustrate the impact of the row numbering or partitioning,
Table \ref{tab:dna_modified} lists timings for two additional
approaches.
\begin{table}[hbpt]
\centering
\begin{tabular}{cc}
\hline
Number of Base Pairs & 9 \\
\hline
\hline
HSC-extension (s) & 8.7 \\
RGF (s) & 7.8 \\
\hline
HSC-extension with {\it customized} partition (s) & 7.8 \\
{\it Customized} RGF with two 2-pairs layers (s) & 11.8 \\
\hline
\end{tabular}
\caption{CPU timings of transmission calculation for DNA molecules
at one energy point.\label{tab:dna_modified}}
\end{table}
The four different partitionings used for these approaches
are depicted on Fig.~\ref{fig:dna9}.
The {\it customized} RGF me\-thod with two 2-pairs layers gathers,
twice, 2-pairs into one layer.
This particular choice of layers indicates the impact
of two blocks of dimension $1000 \times 1000$ on the overall
CPU time, {\it i.e.} an increase in CPU time.
The {\it customized} partition for the HSC-extension
allows a degenerate sub-tree\footnote{Any cluster with 2-pairs
is partitioned according to a degenerate tree, where each parent node
has only one child.} to avoid any 2-pairs cluster.
This {\it customized} partition makes the HSC-extension operate
on blocks of dimension, at most, $500 \times 500$.
This choice results in a lower CPU time, on par with the original
RGF approach.
A similar behavior was observed for other DNA molecules, where METIS
gathers 2-pairs into one cluster.

These numerical experiments suggest that the HSC-exten\-sion,
combined with a multilevel nested dissection, is an efficient
approach even for smaller but denser matrices.
When the graph partitioning is allowed to insert degenerate sub-trees,
the performance is comparable with that of the performance of RGF.

\section{Conclusion}
\label{sec:conc}

In this work, we demonstrate the HSC-extension based NEGF solver as a working methodology for various 3D systems.
The cost analysis for HSC-extension is performed on a cuboid structure.
HSC-extension exhibits operation count of $\mathcal{O}(N^6)$ when simulating cubic device with dimension $N\times N\times N$, whereas a $\mathcal{O}(N^7)$ count is observed for RGF.
We also illustrate various asymptotic costs of HSC-extension when the device has an elongated shape ($N_x, \, N_z \ll N_y$), when the device is flattened ($N_z \ll N_x, \, N_y$), and when a dense self-energy is used to model the open boundary conditions.

The runtime performance of HSC-extension is further investigated for nano-electronic devices of practical interest: graphene-hBN-graphene multilayer heterostructure, silicon nanowire and DNA molecule.
These devices exhibit distinct atomistic sparsity, indicating different computational efficiency for HSC-extension.
The numerical experiments suggest that the HSC-extension exhibits asymptotic runtimes and operation counts proportional to the runtime of the LU-factorization.
For all the nano-electronic devices considered, the HSC-extension becomes faster than the RGF method as the device gets larger.
A 1,000 speed-up is observed for a graphene-hBN-graphene multilayer device with 40,000 atoms.
Since the HSC-extension requires less operations than RGF, these speed-ups will increase as the device gets larger.
A MATLAB version of HSC-extension algorithm is available at \url{https://els.comotion.uw.edu/express_license_technologies/glessuw}.

\begin{acknowledgements}

The authors acknowledge the support by the National Science Foundation
under Grant ECCS 1231927.
S. R. Patil acknowledges the support by UGC-India Grant No. F.5-50/2014(IC).
M. P. Anantram and J. Qi acknowledge the support from the National
Science Foundation under Grant No. 102781 for the DNA part of the paper.
\end{acknowledgements}

\appendix
\section{Complexity Derivation of HSC-extension for 3D Cuboidal Structures}

To analyze the runtime complexity of HSC-extension, we consider a cuboid device with $N_x \times N_y \times N_z$ grid points per direction.

First we discuss the case of a cubic mesh,  {\it i.e.} $N_x = N_y = N_z = N$. The operation count for evaluating diagonal entries in $\mathbf{G}^r$ was discussed by Lin {\it et al.} \cite[section 2.5.3]{Lin2009ab}.
According to their analysis, the operation count grows as $\mathcal{O}(N^6)$.
For the HSC-extension, the complexity for evaluating diagonal entries of $\mathbf{G}^<$ is identical to the complexity for $\mathbf{G}^r$ (as discussed in Hetmaniuk {\it et al.} \cite{hetmaniuk2013nested} for two-dimensional devices).
The overall operation count will grow as $\mathcal{O}(N^6)$.

Next we consider the case of an elongated device, where the numbers of grid points per direction satisfy $N_x = N_z = N \ll N_y$.
The multilevel nested-dissection will identify $N_y/N_z$ subdomains, each discretized with $N \times N\times N$ grid points.
When evaluating diagonal entries in $\mathbf{G}^r$, the operation count for each cubic subdomains will grow as $\mathcal{O}(N^6)$.
The remaining operations will involve dense matrices for the separators of dimension $N_x \times N_z$.
The algebraic operations for one separator will include dense matrix-matrix multiplications and dense matrix inversions, yielding an asymptotic cost $\mathcal{O}(N_x^3N_z^3) = \mathcal{O}(N^6)$.
The number of separators is also $\mathcal{O}(N_y/N)$.
The overall operation count will grow as $\mathcal{O}(N^6)\mathcal{O}(N_y/N) = \mathcal{O}(N^5N_y)$.

Finally we consider the case of a flattened device, where the number of grid points per direction satisfy $N_z \ll N_x = N_y = N$.
As discussed by Lin {\it et al.} \cite[section 2.5.3]{Lin2009ab}, this configuration is similar to a two-dimensional problem with $N \times N$ grid points.
The prefactor will depend on $N_z$.
The HSC algorithm and our extension can proceed as if the device is two-dimensional by replacing scalar algebraic operations with block algebraic operations (each block being of dimension $N_z \times N_z$).
These block operations will cost $\mathcal{O}(N_z^3)$.
So the overall operation count will grow as $\mathcal{O}(N_z^3) \mathcal{O}(N^3) = \mathcal{O}(N_z^3 N^3)$.

\bibliographystyle{spphys}

\begin{thebibliography}{10}

\bibitem{datta2002non}
S.~Datta, in \emph{Electron Devices Meeting, 2002. IEDM'02. International}
  (IEEE, 2002), pp. 703--706

\bibitem{ren2003nanomos}
Z.~Ren, R.~Venugopal, S.~Goasguen, S.~Datta, M.S. Lundstrom, IEEE T. Electron
  Dev. \textbf{50}(9), 1914 (2003)

\bibitem{barker1989theory}
J.R. Barker, J.~Pepin, M.~Finch, M.~Laughton, Solid-state electron.
  \textbf{32}(12), 1155 (1989)

\bibitem{luisier2006quantum}
M.~Luisier, A.~Schenk, W.~Fichtner, J. Appl. Phys. \textbf{100}(4), 043713
  (2006)

\bibitem{asenov2003simulation}
A.~Asenov, A.R. Brown, J.H. Davies, S.~Kaya, G.~Slavcheva, IEEE T. Electron
  Dev. \textbf{50}(9), 1837 (2003)

\bibitem{martinez2007study}
A.~Martinez, K.~Kalna, J.R. Barker, A.~Asenov, Physica E \textbf{37}(1), 168
  (2007)

\bibitem{martinez2006development}
A.~Martinez, J.R. Barker, A.~Asenov, M.~Bescond, A.~Svizhenko, A.~Anantram, in
  \emph{Simulation of Semiconductor Processes and Devices, 2006 International
  Conference on} (IEEE, 2006), pp. 353--356

\bibitem{martinez2007self}
A.~Martinez, M.~Bescond, J.R. Barker, A.~Svizhenko, M.P. Anantram, C.~Millar,
  A.~Asenov, IEEE T. Electron. Dev. \textbf{54}(9), 2213 (2007)

\bibitem{martinez2009comparison}
A.~Martinez, A.R. Brown, A.~Asenov, N.~Seoane, in \emph{Simulation of
  Semiconductor Processes and Devices, 2009. SISPAD'09. International
  Conference on} (IEEE, 2009), pp. 1--4

\bibitem{martinez2010detailed}
A.~Martinez, N.~Seoane, A.R. Brown, A.~Asenov, in \emph{Silicon Nanoelectronics
  Workshop (SNW), 2010} (IEEE, 2010), pp. 1--2

\bibitem{Anantram2008aa}
M.~Anantram, M.~Lundstrom, D.~Nikonov, Proc. IEEE \textbf{96}(9), 1511 (2008)

\bibitem{Svizhenko2002aa}
A.~Svizhenko, M.~Anantram, T.~Govindam, B.~Biegel, R.~Venugopal, J. Appl. Phys.
  \textbf{91}, 2343 (2002)

\bibitem{Lin2009ab}
L.~Lin, J.~Lu, L.~Ying, R.~Car, W.~E, Commun. Math. Sci. \textbf{7}(3), 755
  (2009)

\bibitem{Lin2011ac}
L.~Lin, C.~Yang, J.~Meza, J.~Lu, L.~Ying, W.~E, ACM Trans. Math. Softw.
  \textbf{37}(40) (2011)

\bibitem{Li2008ab}
S.~Li, S.~Ahmed, G.~Klimeck, E.~Darve, J. Comp. Phys. \textbf{227}, 9408 (2008)

\bibitem{Li2011ab}
S.~Li, E.~Darve, J. Comp. Phys. \textbf{231}(4), 1121 (2012)

\bibitem{hetmaniuk2013nested}
U.~Hetmaniuk, Y.~Zhao, M.P. Anantram, Int. J. Numer. Meth. Eng. \textbf{95}(7),
  587 (2013)

\bibitem{Takahashi1973}
K.~Takahashi, J.~Fagan, M.S. Chin, in \emph{Eighth PICA Conference} (1973)

\bibitem{Erisman1975}
A.M. Erisman, W.F. Tinney, Commun. ACM \textbf{18}(3), 177 (1975)

\bibitem{George1973aa}
A.~George, SIAM J. Numer. Anal. \textbf{10}(2), 345 (1973)

\bibitem{Karypis:1998:FHQ:305219.305248}
G.~Karypis, V.~Kumar, SIAM J. Sci. Comput. \textbf{20}(1), 359 (1998)

\bibitem{nissen2011optimized}
A.~Nissen, G.~Kreiss, Commun. Comput. Phys. \textbf{9}, 147 (2011)

\bibitem{matlab}
{The MathWorks Inc.}, \emph{{MATLAB Release 2011b}}.
\newblock Natick, Massachusetts, United States (2011)

\bibitem{umfpack}
T.~Davis, \emph{Direct Methods for Sparse Linear Systems} (SIAM, 2006)

\bibitem{gastro2009}
A.H. Castro~Neto, F.~Guinea, N.M.R. Peres, K.S. Novoselov, A.K. Geim, Rev. Mod.
  Phys. \textbf{81}(1), 109 (2009)

\bibitem{britnell2013resonant}
L.~Britnell, R.V. Gorbachev, A.K. Geim, L.A. Ponomarenko, A.~Mishchenko, M.T.
  Greenaway, T.M. Fromhold, K.S. Novoselov, L.~Eaves, Nat. Comm. \textbf{4},
  1794 (2013)

\bibitem{zhao2015negative}
Y.~Zhao, Z.~Wan, X.~Xu, S.R. Patil, U.~Hetmaniuk, M.P. Anantram, Scientific
  reports \textbf{5} (2015)

\bibitem{cui2001functional}
Y.~Cui, C.M. Lieber, Science \textbf{291}(5505), 851 (2001)

\bibitem{luisier2006atomistic}
M.~Luisier, A.~Schenk, W.~Fichtner, G.~Klimeck, Phys. Rev. B \textbf{74}(20),
  205323 (2006)

\bibitem{Gohler18022011}
B.~G\"{o}hler, V.~Hamelbeck, T.Z. Markus, M.~Kettner, G.F. Hanne, Z.~Vager,
  R.~Naaman, H.~Zacharias, Science \textbf{331}(6019), 894 (2011)

\bibitem{postma2010rapid}
H.W.C. Postma, Nano Lett. \textbf{10}(2), 420 (2010)

\bibitem{tsutsui2011electrical}
M.~Tsutsui, K.~Matsubara, T.~Ohshiro, M.~Furuhashi, M.~Taniguchi, T.~Kawai, J.
  Am. Chem. Soc. \textbf{133}(23), 9124 (2011)

\bibitem{g09}
M.J. Frisch, G.W. Trucks, H.B. Schlegel, G.E. Scuseria, M.A. Robb, J.R.
  Cheeseman, G.~Scalmani, V.~Barone, B.~Mennucci, G.A. Petersson, H.~Nakatsuji,
  M.~Caricato, X.~Li, H.P. Hratchian, A.F. Izmaylov, J.~Bloino, G.~Zheng, J.L.
  Sonnenberg, M.~Hada, M.~Ehara, K.~Toyota, R.~Fukuda, J.~Hasegawa, M.~Ishida,
  T.~Nakajima, Y.~Honda, O.~Kitao, H.~Nakai, T.~Vreven, J.A. Montgomery, {Jr.},
  J.E. Peralta, F.~Ogliaro, M.~Bearpark, J.J. Heyd, E.~Brothers, K.N. Kudin,
  V.N. Staroverov, R.~Kobayashi, J.~Normand, K.~Raghavachari, A.~Rendell, J.C.
  Burant, S.S. Iyengar, J.~Tomasi, M.~Cossi, N.~Rega, J.M. Millam, M.~Klene,
  J.E. Knox, J.B. Cross, V.~Bakken, C.~Adamo, J.~Jaramillo, R.~Gomperts, R.E.
  Stratmann, O.~Yazyev, A.J. Austin, R.~Cammi, C.~Pomelli, J.W. Ochterski, R.L.
  Martin, K.~Morokuma, V.G. Zakrzewski, G.A. Voth, P.~Salvador, J.J.
  Dannenberg, S.~Dapprich, A.D. Daniels, O.~Farkas, J.B. Foresman, J.V. Ortiz,
  J.~Cioslowski, D.J. Fox.
\newblock Gaussian~09 {R}evision {A}.1.
\newblock Gaussian Inc. Wallingford CT 2009

\bibitem{qi2013unified}
J.~Qi, N.~Edirisinghe, M.G. Rabbani, M.P. Anantram, Phys. Rev. B
  \textbf{87}(8), 085404 (2013)

\end{thebibliography}

\end{document}